\documentclass[%
 reprint,a4paper,
superscriptaddress,
 amsmath,amssymb,
 aip,jcp,
]{revtex4-1}

\usepackage{pdfpages}
\usepackage{pgffor}
\usepackage{graphicx}%
\usepackage{bm}%
\usepackage[normalem]{ulem}
\usepackage{xcolor}
\usepackage{amsmath,amssymb,amsfonts,mathtools,cancel}

\definecolor{DragonGreen}{RGB}{0,126,48}
\definecolor{Brownie}{RGB}{97,16,9}

\usepackage[paperwidth=210mm,
            paperheight=297mm,
            left=20mm,
            top=10mm,
            textwidth=170mm,
            marginparsep=3mm,
            marginparwidth=30mm,
            textheight=730pt,
            footskip=50pt,portrait]
           {geometry}

\newcommand{\avg}[1]{\ensuremath{\langle{#1}\rangle}}

\newcommand{\ns}{\ensuremath{{n_{\text{s}}}}} %
\newcommand{\nl}{\ensuremath{n_{\text{l}}}}
\newcommand{\ntot}{\ensuremath{n_{\text{tot}}}}
\newcommand{\nsites}{\ensuremath{N_{\text{s}}}}
\newcommand{\ncut}{\ensuremath{n_{\text{cut}}}}

\newcommand{\phis}{\ensuremath{\phi_{\text{s}}}} %
\newcommand{\phil}{\ensuremath{\phi_{\text{l}}}}
\newcommand{\Phil}{\ensuremath{\Phi_{\text{l}}}}

\newcommand{\cond}[2]{\ensuremath{\left(#1\middle|#2\right)}}
\newcommand{\D}{\ensuremath{\mathrm{d}}}
\newcommand{\barns}{\ensuremath{\bar{n}_\mathrm{s}}}
\newcommand{\rhosl}{\ensuremath{\rho_\mathrm{sl}}}

\begin{document}

\preprint{APS/123-QED}

\title{
Bridging the gap between atomistic 
and \\ macroscopic models of
homogeneous nucleation
}%

\author{Bingqing Cheng}
\email{bingqing.cheng@epfl.ch}
 \affiliation{Laboratory of Computational Science and Modeling, Institute of Materials, {\'E}cole Polytechnique F{\'e}d{\'e}rale de Lausanne, 1015 Lausanne, Switzerland}%

\author{Michele Ceriotti}
\affiliation{Laboratory of Computational Science and Modeling, Institute of Materials, {\'E}cole Polytechnique F{\'e}d{\'e}rale de Lausanne, 1015 Lausanne, Switzerland}%

\date{\today}%

\begin{abstract}

Macroscopic theories of nucleation such as 
classical nucleation theory envision that
clusters of the bulk stable phase form 
inside the bulk metastable phase.
Molecular dynamics simulations are often
used to elucidate 
nucleation mechanisms, 
by capturing 
the microscopic configurations of 
all the atoms.
In this paper,
we introduce a thermodynamic model
that links macroscopic theories and 
atomic-scale simulations 
and thus provide a simple and elegant
framework for testing the limits of 
classical nucleation theory.

\end{abstract}
\keywords{nucleation, atomistic simulation, thermodynamics, statistical mechanics}%
\maketitle

\begin{figure}[htbp]
\includegraphics[width=1.0\columnwidth]{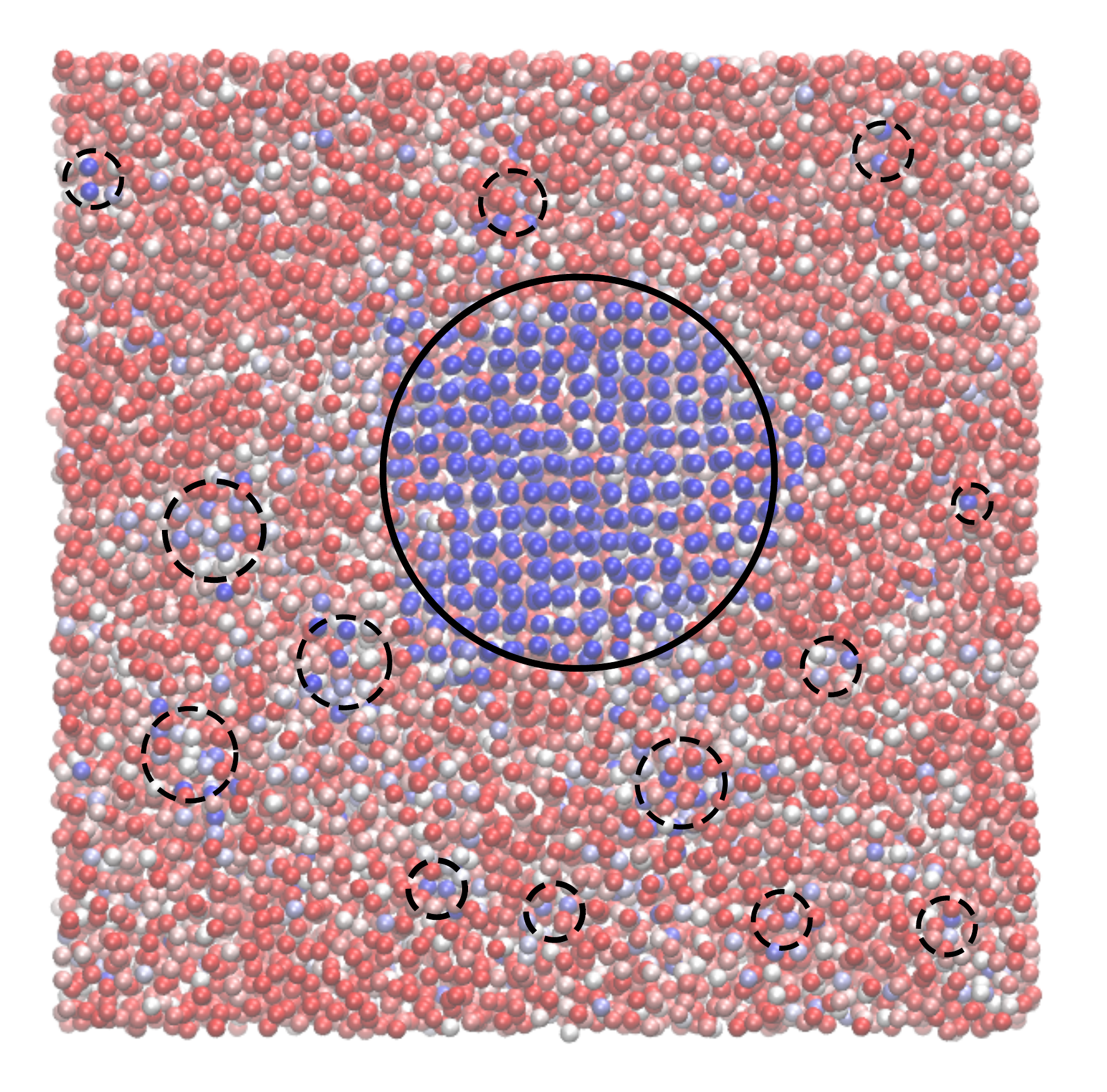}
\caption{
A snapshot of an under-cooled liquid system that has a large solid cluster (solid circle),
and many smaller clusters (dashed circles).
Atoms are colored based on the value of a local order parameter so
solid-like atoms are colored in blue,
while liquid-like atoms are colored in red.
Details on the underlying simulation are
given in Section~\ref{sec:applications}.
}
\label{fig:p1}
\end{figure}

Nucleation is a key step in bulk phase 
transitions~\cite{oxtoby1992homogeneous,schmelzer2005nucleation,yi2012molecular,sosso2016crystal}.
This process plays a crucial role in natural
phenomena and in technological applications,
from the formation of clouds~\cite{sosso2016crystal} 
to self-assembly\cite{jonkheijm2006probing},
and from casting to the growth of thin films~\cite{Boettinger200043,venables1984nucleation}.
One of the simplest models to 
rationalize nucleation is classical 
nucleation theory (CNT), which 
assumes that the stable phase 
forms by accretion of nanoscopic
nuclei. 
These clusters are unstable when they
are smaller than a critical size 
$n^\star$, and at any given time
the metastable phase contains multiple 
sub-critical clusters of the 
stable phase (Figure~\ref{fig:p1}).
The average number of 
clusters containing $n$ atoms in the system is given by:
\begin{equation}
    \avg{p_n} \propto \exp(-\beta G(n)),
    \label{eq:P}
\end{equation}
where $\beta = 1/k_B T$, and $G(n)$ is the free 
energy excess associated with a single cluster of 
size $n$.
In the context of homogeneous nucleation,
CNT further assumes that 
$G(n)$ can be expressed as the sum of a bulk 
and a surface term, i.e.
\begin{equation}
    G(n)=\mu n + \sigma v^{\frac{2}{3}} n^{\frac{2}{3}},
    \label{eq:cnt}
\end{equation}
where $\mu$ is the chemical potential difference 
between the stable and the metastable phases,
$\sigma$ is the effective interfacial free energy,
and $v$ is the molar volume of the 
bulk stable phase.

Investigating experimentally the nature and behavior of the 
unstable subcritical nuclei is extremely difficult. 
Therefore, in the last
two decades, a considerable number of 
atomistic simulation studies have been devoted to 
investigating homogeneous nucleation,
especially to
verifying the accuracy of the CNT model
~\cite{ten1996numerical,ten1999homogeneous,auer2001prediction,moroni2005interplay,trudu2006freezing,maibaum2008phase,lechner2011role,prestipino2012systematic,yi2012molecular,salvalaglio2015molecular,mccarty2016bespoke,piaggi2016fdreact16,Lifanov2016}.
Some of these studies have found a good agreement 
between the CNT prediction in Eqn.~\eqref{eq:cnt} 
and the free energy profile for a cluster $G(n)$ that 
was computed from 
simulations~\cite{auer2001prediction,piaggi2016fdreact16}.  Others, meanwhile, have shown significant systematic 
differences between the two~\cite{prestipino2012systematic}. 

\begin{figure}[hpbt]
\includegraphics[width=1.0\columnwidth]{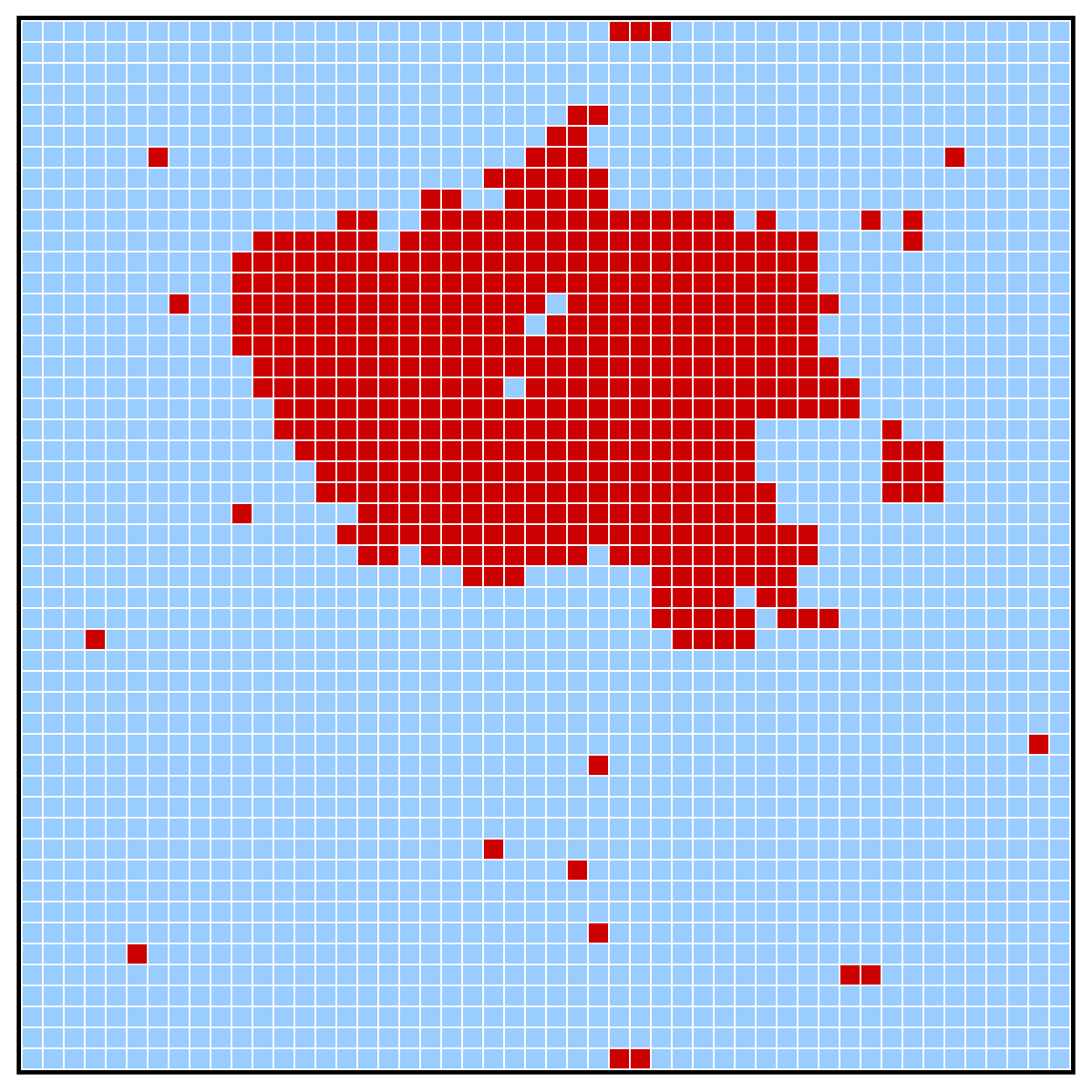}
\caption{
A snapshot of a two-dimensional  square-lattice Ising model undergoing a transition between ``up'' (blue) and ``down'' (red) ferromagnetic phases. 
Details on the underlying simulation are
given in Section~\ref{sec:applications}.
}
\label{fig:ising-nucleus}
\end{figure}

{While there are physical 
reasons why a system might deviate 
from the predictions of CNT\cite{prestipino2012systematic}, 
one should also consider that
there are practical difficulties in applying
an expression such as 
Eqn.~\eqref{eq:cnt} -- that
was designed to be valid at
the macroscopic limit where
phases are well-defined and interfaces can be regarded as perfectly sharp -- to
atomic-scale simulations.}
{When analysing an atomistic
model, one typically
proceeds by first selecting an
arbitrary order parameter that is 
able to distinguish between the 
atoms in each of the two phases.  
The atoms that are thus identified 
as being part of the more stable 
phase are then grouped into 
clusters~\cite{auer2001prediction,moroni2005interplay}.
These heuristic procedures} make
the definition of the clusters 
size $n$ and the associated free energy 
profile $G(n)$ ambiguous.
More importantly, however, there is a 
conceptual gap in assuming that
fluctuations in the metastable phase
involving a few atoms should be regarded
as a nucleus of a stable phase that
is only defined in the thermodynamic limit.

In this paper we address the problem of how to 
reconcile the picture emerging from simulation with macroscopic theories
of nucleation.
To achieve this, we first investigate a
multiple cluster model that we use as a proxy for an idealized atomistic system.
Then, we develop a thermodynamic framework
that is consistent with the multiple cluster model, 
requires fewer assumptions,
and is fully applicable to the atomistic 
systems simulated in molecular dynamics
or Monte Carlo studies. 
For the sake of clarity, we will develop our theoretical framework
making reference to the case of solidification from the melt  (Figure~\ref{fig:p1}),
but our results  are general, and we present an application
to a two-dimensional Ising model (Figure~\ref{fig:ising-nucleus}) to 
demonstrate that it can be
applied to all sorts of activated phase transition processes. 

\section{A thermodynamic model
of atomic-scale nucleation events}
\subsection{An idealized multiple-cluster model}
\begin{figure}
\includegraphics[width=0.5\textwidth]{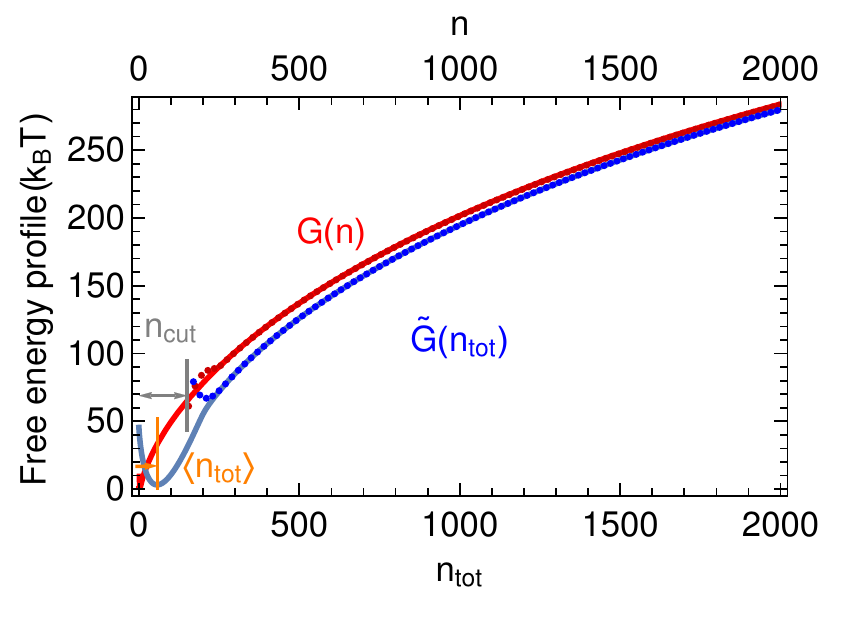}
\caption{
The red line represents the free energy profile $G(n)$ of a single cluster. 
The blue line shows the exact $\tilde{G}(\ntot)$ of the atomistic system with multiple clusters.
The red dots and the blue dots indicate 
$G(n)$ and $\tilde{G}(\ntot)$ that are approximated using Eq.~\eqref{eq:reverse}, respectively.
The grey and yellow vertical lines indicate $\ncut$ and
$\avg{\ntot}$, respectively.
}
\label{fig:p2}
\end{figure}

We start by taking an idealized model of a 
metastable bulk liquid phase, 
in which all the solid clusters can be identified 
unambiguously.  We then further assume that
the interactions between clusters are insignificant 
(e.g. negligible volume exclusion).
We use the symbol $p_n$ to denote the number of solid 
clusters containing $n$ atoms.  The total number of 
solid atoms in the system
is thus just $\ntot=\sum_{n=1}^{\infty} n p_n$.
If the average population of cluster sizes 
follows Eqn.~\eqref{eq:P}, 
the probability distribution for the cluster populations $P(n,p_n)$ can be approximated using a Poisson distribution, i.e.
{
\begin{equation}
\begin{split}
    P(n,p_n) = & \lambda(n)^{p_n} e^{-\lambda(n)}/p_n!~,\\
    \lambda(n) = \avg{p_n} = & \nsites e^{-\beta G(n)},
\end{split}
\label{eq:count}
\end{equation}
where $G(n)$ is the free energy of a single cluster of size $n$,
and $\nsites$ is the number of nucleation sites.
$\nsites$ is proportional to the total number of particles
in the system, and guarantees the 
appropriate scaling with system size.
$G(n)$ can take any form as long as
it is a monotonically increasing function for $n$ smaller than the size of the critical nucleus,
but for this idealized system we use Eqn.~\eqref{eq:cnt} 
with the parameters reported in the Supplementary Material.  The free energy profile for this equation with these parameters is shown in red in Figure~\ref{fig:p2}.
}

 As we have assumed 
that the cluster size distribution follows Eqn.~\eqref{eq:count},
 we can derive the following expression for the free energy profile of the whole system as a function of the \emph{total} number of solid atoms, $\tilde{G}(\ntot)$,
\begin{multline}
        e^{-\beta \tilde{G}(\ntot)} = \\
    \sum_{p_1=0}^{\infty} \sum_{p_2=0}^{\infty} \ldots\sum_{p_{n^\star}=0}^{\infty} 
    \delta\left(\sum_{n=1}^{\infty} n p_n -\ntot\right)
    \prod_{n=1}^{n^\star} P(n,p_n),
    \label{eq:gtot}
\end{multline}
by explicitly enumerating all the possible combinations of cluster sizes 
that result in the same $\ntot$.
Here $n^\star$ is the size of the critical nucleus,
which is taken as an upper bound for 
the cluster size, in order to restrict the analysis
to the range of metastability of the liquid~\cite{maibaum2008phase}.
We computed $\tilde{G}(\ntot)$ 
analytically using Eqn.~\eqref{eq:gtot}, 
and plotted the result in blue in Figure~\ref{fig:p2}.
It is important to note that $\tilde{G}(\ntot)$ does not 
explicitly depend on the 
size composition of all the clusters. Defining this 
quantity is thus an important step towards the formulation of a 
macroscopic view of nucleation that is reliant on extensive quantities 
calculated over the whole system.

 Figure~\ref{fig:p2} shows that
 the liquid contains $\avg{\ntot}$ solid atoms on average.
By writing out explicitly the cluster size composition, we noticed that 
when $\ntot \lesssim \avg{\ntot}$ the most probable configuration for the system was composed
 of several small clusters.
However, as $\ntot$ gets larger,
it typically contained 
one large solid cluster accompanied by
many smaller ones.
As we discuss in detail in the Supplementary Material,
one can thus define a cutoff size $\ncut$,
such that for $\ntot\gg\ncut$  it is orders of magnitude 
more likely to have precisely one cluster with size $n>\ncut$ than
to have several or none of such large clusters.
At $\ntot\gg\ncut$,
the largest cluster can be interpreted as 
a standalone solid cluster with $n$ 
atoms, associated with a probability
$P(n,1)$. The rest of clusters in the background
can be treated as a separate 
bulk liquid system that follows the same 
distributions (Eqn.~\eqref{eq:count}) as the original whole system.
Under such treatment, at $\ntot\gg\ncut$ the
expression for $\tilde{G}(\ntot)$ can be
simplified tremendously as
{
\begin{equation}
\exp(-\beta \tilde{G}(\ntot)) = \sum_{n = \ncut}^{\ntot} 
\nsites e^{-\beta G(n)}
e^{-\beta \tilde{G}(\ntot - n)}
    \label{eq:reverse},
\end{equation}
considering also that
 $P(n,1)\approx \nsites\lambda(n)$ for large $n$.}
The blue dots in Figure~\ref{fig:p2} correspond to
the approximate $\tilde{G}(\ntot)$ that can be computed using Eqn.~\eqref{eq:reverse}.  These points overlap perfectly with the exact values at $\ntot\gtrsim \ncut+\avg{\ntot}$.
What is more, 
Eqn.~\eqref{eq:reverse} suggests that
there is a one-to-one mapping between
the $\tilde{G}(\ntot)$ for the whole system
and the $G(n)$ for a single cluster with $n\gtrsim \ncut$.
As such,
Eq.~\eqref{eq:reverse} allows us to calculate $G(n)$ from a knowledge of $\tilde{G}(\ntot)$, 
without any information on the sizes of individual clusters in each snapshot.
The $G(n)$ that is reconstructed from Eq.~\eqref{eq:reverse} using a fixed-point numerical scheme is indicated using red dots in Figure~\ref{fig:p2},
and overlaps perfectly with the exact $G(n)$.

\subsection{A probabilistic definition of Gibbs dividing surfaces}

Our discussion thus far demonstrates that
the average distribution of cluster sizes can be 
extracted from the distribution of the 
total number of atoms assigned to the stable phase. 
Unfortunately, in actual atomistic simulations it is impossible
to assign individual atoms or molecules to either of the
two phases without additional empirical assumptions. 
In what follows, we will therefore {use the concept of Gibbs dividing surface, and } introduce a thermodynamic approach that 
draws a
connection between an atomistic and a macroscopic 
description of homogeneous nucleation.

{
A Gibbs dividing surface is defined to be an infinitely
thin geometrical surface that is 
sensibly coincident with the physical surface of discontinuity~\cite{gibbs1928collected,tolman1948consideration}.
The surface is meant to be an idealization
of the transition region 
between the two phases, and one should choose,
whenever possible, a geometry that is consistent
with the boundary conditions and the symmetry 
of the problem.
The precise position and shape of the 
Gibbs dividing surface are important, for instance
when one needs to determine its area~\cite{gibbs1928collected,tolman1948consideration}.
When instead one
only needs to define the extent of the two
bulk phases, what matters most is that 
the surface divides the system into a solid part that has $\ns$ atoms and a liquid part that has $\nl$ atoms, with no atom assigned to
the interface.
It is then useful to construct a reference system,
in which the solid part and the liquid part maintain their bulk properties up to the dividing surface.
With a dividing surface in place, the free energy and the properties of any two-phase system can be 
naturally decomposed into a term
corresponding to the reference bulk system and an excess term associated with the interface.

To remove the degree of freedom associated with the choice of a dividing surface,
it is customary to select a surface such that
there is no surface excess of a certain extensive quantity $\Phi$,
i.e. such that the real system and the reference system exhibit the same value for the chosen extensive quantity.
This extensive quantity could be the volume 
occupied by that region, its internal
energy or its entropy for example.
More often than not, it is also convenient
to use an extensive
order parameter $\Phi = \sum_i \phi_i$,
where the atomic order parameter $\phi_i$ is calculated based on the local environment of each of the particles in the system.
The zero surface excess condition can be schematically expressed as
\begin{equation}
    \Phi_{\text{sl}}(\ns,\nl) \equiv \Phi_{\text{ref}}(\ns,\nl),
    \label{eq:gibbs}
\end{equation}
where $\Phi_{\text{sl}}(\ns,\nl)$ and $\Phi_{\text{ref}}(\ns,\nl)$ stands
for the values of the extensive quantity $\Phi$ of the real solid-liquid system and the reference system that both comprise $\ns$ solid atoms and $\nl$ liquid atoms.

Now consider a solid-liquid system comprising a total of $N$ atoms.
 In our previous work~\cite{cheng2015solid} we argued 
 that an ideal reference system for any microstate of 
 such a system can be constructed based 
 on the instantaneous value of $\Phi$ in that microstate.
This reference system comprises a bulk solid that has $\ns$ atoms,
and a bulk liquid that has $N-\ns$ atoms.
To find the value of $\ns$ in this system one simply applies the deterministic mapping $\Phi_\text{ref}(\ns,\nl)=\phis \ns + \phil (N-\ns)$ where $\phis$ and $\phil$ are the average value for the order parameter of each atom in the solid and liquid respectively.
This mapping corresponds to a Gibbs dividing surface between the two phases
that has zero excess for the extensive variable $\Phi$.

However, it is important to realize that an instantaneous extensive quantity
of a finite piece of bulk solid or bulk liquid can fluctuate even at fixed thermodynamic conditions.
As such, the extensive quantity $\Phi$ of a reference system 
that has a bulk solid part with $\ns$ atoms and a bulk liquid part with $N-\ns$ atoms also have fluctuations.
In what follows, we will describe a probabilistic framework that
takes into account the fluctuations when determining
the dividing surface.

First, consider an unbiased simulation of the 
bulk solid phase. Then, select contiguous portions 
of the solid of varying size, and determine for each
case the number of atoms $n$ contained in the 
region and the value of the extensive quantity $\Phi$.
By computing the histogram of these quantities one
can estimate
\begin{equation}
    \rho_s\cond{\Phi}{n} = \int  \delta(\Phi(\Omega) -\Phi) \D\Omega /\int \D\Omega,
    \label{eq:rhophi}
\end{equation}
where $\Omega$ denotes a possible microstate for the $n$ atoms of that region, distributed with a probability consistent with the thermodynamic conditions, and
$\Phi(\Omega)$ is the value of $\Phi$ for that microstate.
This distribution, $\rho_s\cond{\Phi}{n}$, can be regarded as the conditional probability for observing 
$\Phi$ in a system consists of a bulk solid region containing $n$ atoms.
Since this solid region should mimic the bulk solid part in the reference system, 
strictly speaking its shape should be delimited by the the Gibbs dividing surface and the boundaries of the reference system.
It is worth noting, however, that in the cases we considered in this study the shape of the solid region has little impact on $\rho_s\cond{\Phi}{n}$,
as long as a compact shape is chosen.
An analogous distribution $\rho_l\cond{\Phi}{n}$ 
can be derived for a bulk liquid region that contains $n$ atoms
and also has a shape determined by the dividing surface and the boundaries of the reference system.

Contrary to the multiple cluster model discussed
above, ``solid'' and ``liquid'' in this case indicate well-defined thermodynamic
states.
The bulk solid state encompasses all the possible configurations for a system of solid,
which can contain point defects, other crystal defects, and even small molten pools.
Similarly, a bulk liquid comprises local crystalline orderings and sub-critical solid clusters.

The finite width of the distributions for $\rho_s\cond{\Phi}{n}$ and $\rho_l\cond{\Phi}{n}$ in Eqn.~\eqref{eq:rhophi} ensures that
the value of $\Phi$ for a given microstate cannot
be used to determine the composition of a reference system with absolute certainty.
Instead, 
we can compute the distribution of $\Phi$ for a
reference system composed of $\ns$ solid atoms and $\nl$ liquid atoms using 
\begin{equation}
    \rho_{\text{ref}}\cond{\Phi}{\ns,\nl}=\int \D\varphi \rho_s\cond{\varphi}{\ns} \rho_l\cond{\Phi-\varphi}{\nl}.
    \label{eq:rhosl}
\end{equation}
According to the concept of the Gibbs dividing surface,
if the reference and the actual system both have $\ns$ solid atoms and 
$\nl$ liquid atoms, they should also both have the same distribution for $\Phi$. In other words, 
the zero-excess condition in Eqn.~\eqref{eq:gibbs} take 
a probabilistic form,
\begin{equation}
    \rhosl\cond{\Phi}{\ns,\nl} \equiv \rho_{\text{ref}}\cond{\Phi}{\ns,\nl}.
    \label{eq:gibbsprob}
\end{equation}
Only when $\rho_s\cond{\Phi}{n}$ and $\rho_l\cond{\Phi}{n}$
are both $\delta$ functions at any given $n$, Eqn.~\eqref{eq:rhosl} is reduced to a deterministic mapping
between $\Phi$ and $\ns$ analogous to that
introduced in Ref.~\cite{cheng2015solid}. 
}

\subsection{Obtaining cluster-size free energies 
from an extensive order parameter}
Let us now describe how to
extract the free energy profile for a solid cluster
from atomistic simulations of undercooled liquid.
In simulations, the values of $\Phi$ can be easily computed for every microstate,
so the associated free energy $\tilde{G}(\Phi)$ can be directly 
obtained from biased or unbiased molecular dynamics simulations. 
Since the atomistic simulations are constructed so that they sufficiently sample all configurations in the undercooled liquid, the computed free energy profile $\tilde{G}(\Phi)$
directly characterizes the distribution of $\Phi$ in the liquid, i.e.
$\rho_l\cond{\Phi}{N} = \exp(-\beta\tilde{G}(\Phi))$.
On the other hand, the bulk liquid sampled in simulations can have configurations that contain
sub-critical nuclei of large sizes,
such as the one illustrated in Figure~\ref{fig:p1}.
Those configurations can have a value of $\Phi$ approaching those typically encountered for a 
solid sample. 
As we have discussed in the multiple cluster model, configurations that
contain a large number of solid-like atoms are overwhelmingly likely
to comprise one and only one cluster of size larger than $\ncut$
and a liquid-like background.
By a similar logic, a configuration with a value of $\Phi$ that has enough solid-like characteristics
can be interpreted
as a single solid cluster larger than $\ncut$
and the surrounding liquid.

Consider a single solid cluster that has $\ns > \ncut$ atoms together with a liquid background of $N-\ns$ atoms.
The values of the extensive quantity for this combination of phases
follow the distribution $\rhosl\cond{\Phi}{\ns,N-\ns}$.
Inside the undercooled bulk liquid,
the average population for solid clusters of size $\ns$ can be expressed as {$\avg{p_{n_s}}=N_s
\exp(-\beta G(\ns))$, where $N_s$ is the number of nucleation sites that can often be considered as the number of atoms or molecules or lattice sites in homogeneous nucleation,
and $\exp(-\beta G(\ns))$ the probability that a nucleus of size $\ns$ has grown around a nucleation site in the metastable liquid.
In other words, $G(\ns)$ represents the free energy excess associated with a solid cluster that has $\ns$ atoms relative to the bulk liquid.
Notice also that for $\ns > \ncut$, the average population  $N_s
\exp(-\beta G(\ns))$ is also the probability of observing a solid cluster of size $\ns$ in the bulk liquid system.
}
Based on these considerations,
{and using the law of total probability,}
the probability 
distribution for $\Phi$ in such systems
follows
{
\begin{equation}
    e^{-\beta\tilde{G}(\Phi)}=  \int_{\ncut}^{n^{\star}} \!\!\!\D \ns
       \rhosl\cond{\Phi}{\ns,N-\ns} N_s e^{-\beta G(\ns)}.
   \label{eq:gphigns}
\end{equation}
}
This expression is valid for 
values of $\Phi$ that satisfy $\rhosl\cond{\Phi}{n,N-n} \approx 0$ for all $n < \ncut$,
so that the system can be considered to have a single cluster of size larger than $\ncut$.

In principle, $G(\ns)$ can be determined from Eqn.~\eqref{eq:gphigns},
as both $\tilde{G}(\Phi))$ and $\rhosl\cond{\Phi}{\ns,N-\ns}$
can be computed from simulations.
However, to avoid the numerical instabilities in the direct deconvolution process, we cast the problem as a
fixed-point iteration.
The average $\Phi$ value for a system containing  $\ns$ solid atoms,
$\bar{\Phi}(\ns)=\int \D \Phi  \rhosl\cond{\Phi}{\ns,N-\ns} \Phi$,
follows a monotonic relation with $\ns$, as shown in the Supplementary Material.
One can invert this relation and obtain a value $\barns(\Phi)$ at each $\Phi$ such that $\Phi=\bar{\Phi}(\barns(\Phi))$. 
More generally, after some simple manipulations, we can rewrite Eqn.~\eqref{eq:gphigns} as:
{
\begin{multline}
\tilde{G}(\Phi)= 
G(\barns(\Phi)) -\frac{1}{\beta} \log N_s\\
-\frac{1}{\beta}\log \int_{\ncut}^{n^\star} \!\!\!\D n \rhosl\cond{\Phi}{n,N-n} e^{-\beta [G(n)-G(\barns(\Phi))]}
.
\end{multline}
}
This equation can be rearranged, exploiting the inversion between $\barns$ and $\Phi$, into a self-consistency condition on $G(\ns)$
{
\begin{multline}
G(\ns)= \tilde{G}(\bar{\Phi}(\ns)) +\frac{1}{\beta} \log N_s\\
+\dfrac{1}{\beta}\log \int_{\ncut}^{n^\star} \!\!\!\D n
\rhosl\cond{\bar{\Phi}(\ns)}{n,N-n} e^{-\beta[ G(n)-G(\ns)]}
,\label{eq:iterative}
\end{multline}
}
which can be solved iteratively 
starting from the initial guess
$G_0(\ns)=\tilde{G}(\bar{\Phi}(\ns))$,
and plugging the old guess
onto the right-hand side to obtain 
a new estimate at each iteration. 
Upon convergence, $G(\ns)$ is an 
estimate of the free energy for a solid cluster containing $\ns$ atoms relative to the bulk liquid.

It is worth stressing that
the cluster size $\ns$ and the 
associated free energy $G(\ns)$ in Eqn.~\eqref{eq:gphigns}
are still dependent on the choice of $\Phi$,
because the reference system is defined 
based on a Gibbs dividing 
surface that has zero excess
for the extensive quantity $\Phi$.
Due to the diffuse nature of the physical interface,
a different choice for the extensive variable can
result in a different location of the Gibbs dividing surface and a different reference system.
However, as extensively discussed in our previous work,
as long as one uses one extensive quantity and its associated reference system consistently throughout the analysis,
no ambiguity will arise
in the value of the nucleation barrier~\cite{cheng2015solid}.

\section{Applications of the 
fluctuating reference framework
\label{sec:applications}
} 

\subsection{Solidification of a Lennard-Jones system}
\begin{figure}
\includegraphics[width=0.5\textwidth]{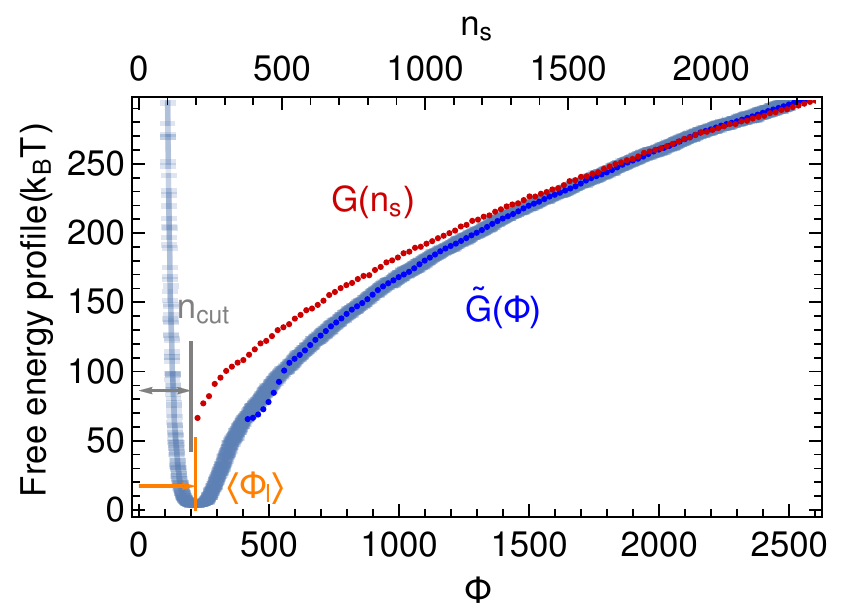}
\caption{
The solid blue line is the free energy profile $\tilde{G}(\Phi)$,
with statistical errors indicated by the error bars.
The red and the blue dots indicate the reconstructed 
curves for $G(\ns)$ and $\tilde{G}(\Phi)$,
respectively.
The grey and yellow vertical lines indicate $\ncut$ and
the average extensive quantity $\avg{\Phil}$, respectively.
}
\label{fig:p3}
\end{figure}

To demonstrate how this thermodynamic framework
can be applied to an atomistic simulation
of a phase transition,
{we simulated the processes of homogeneous solidification for 
a Lennard-Jones system of 23328 atoms at $T=0.58$~\cite{davidchack2003direct,angi+10prb,benjamin2014crystal} -- corresponding to a moderate undercooling relative to the melting temperature of  $T_m=0.6185$.~\cite{davidchack2003direct,angi+10prb,cheng2015solid}}
We performed 12 independent biased sampling runs using the well-tempered metadynamics 
protocol with adaptive Gaussians ~\cite{plim95jcp,barducci2008well,branduardi2012metadynamics,tribello2014plumed}.
The sample input files can be found in the Supplementary Material.
We used a collective variable $\Phi=\sum_i S(\kappa(i))$ in the biased simulations,
where $S(\kappa(i))$ is the local atomic order parameter for atom $i$, as described in Ref.~\cite{cheng2015solid}.
The solid blue line in Figure~\ref{fig:p3} indicates the free energy profile $\tilde{G}(\Phi)$ that was obtained by re-weighting the trajectories.
{Assuming the Gibbs dividing surface separating the solid cluster and the bulk liquid has a spherical shape,
we also computed the two probability distributions $\rho_s\cond{\Phi}{n}$ and $\rho_l\cond{\Phi}{n}$ from unbiased simulations of the bulk phases as shown in the Supplementary Material.}

The snapshot in Figure~\ref{fig:p1} is taken from one of the biased runs, with
each atom colored according to the value of $S(\kappa(i))$. 
Analyzing the population of cluster sizes in this snapshot requires a man-made choice of a cutoff value for 
$S(\kappa)$, and a complex procedure to
identify adjacent groups of solid atoms.  
Instead, by applying the 
thermodynamic model introduced above,
we can simply characterize the behavior
of $\Phi$ in the solid and the liquid
phases, and use that knowledge to convert
$\tilde{G}(\Phi)$ of the whole system into
$G(\ns)$,
using the iterative expression in
Eqn.~\eqref{eq:iterative}.
{Here we also assume the total number of atoms in the system $N$ to be the number of nucleation sites $N_s$, {although any other choice would
simply amount to a vertical shift of
the free-energy curve.}}
This curve of $G(\ns)$
plotted as the red dots in
Figure~\ref{fig:p3},
corresponds to the free energy for a single cluster {relative to the bulk liquid.}
In order to demonstrate the 
convergence of the conversion, we used the computed $G(\ns)$
to reconstruct the free energy profile
$\tilde{G}(\Phi)$
using Eqn.~\eqref{eq:gphigns}.
As shown in Figure~\ref{fig:p3},
the reconstructed $\tilde{G}(\Phi)$
is indistinguishable from that
obtained directly from the 
simulation.

As suggested by the
many similarities between 
Figure~\ref{fig:p3} and Figure~\ref{fig:p2},
the multiple cluster model and the thermodynamic model 
are very 
closely related. 
In the Supplementary Material we show that
under a few additional assumptions
Eqn.~\eqref{eq:gphigns} is exactly
the same as Eqn.~\eqref{eq:reverse}, with
$\Phi$ taking the role of $\ntot$.
Eqn.~\eqref{eq:gphigns}
serves a dual purpose: it 
converts the extensive 
quantity $\Phi$ into an 
estimate for the overall 
solid fraction and it singles
out the free-energy excess for 
the largest cluster from the
fluctuations of the background
liquid.
\subsection{Nucleation in a two-dimensional Ising Model} 

{
In order to demonstrate the
general applicability of our 
thermodynamic framework, 
we discuss in this section its application to
the homogeneous nucleation of a 
two-dimensional Ising model, in the 
absence of external magnetic field.
The model is described by the usual first-neighbour Heisenberg Hamiltonian
\begin{equation}
    H = -J\sum_{\avg{i,j}} s_i s_j,
\end{equation}
where $J=1$ is the coupling constant,
the spin $s_i$ at site $i$ is either up (+1) or down (-1),
and the sum extends over all its nearest neighbors in the lattice.
We used a periodic square lattice with side $L=25$,
and performed a Monte Carlo simulation with biased sampling~\cite{torr-vall99jcp} at the temperature $T=1.5$, well below the 
critical temperature $T_c=2.269$.
We started the simulation with all spins down,
and also restricted the sampling to the states with negative total magnetization.

The snapshot in Figure~\ref{fig:ising-nucleus}
(from a simulation with $L=50$) shows 
a large cluster with positive magnetization embedded in the phase with spins down.
Note that 
one can observe spontaneous fluctuations
of opposite spins not only in the 
negatively-magnetized background,
but also inside the large 
nucleating cluster, underscoring the 
ambiguity in defining cluster sizes
by counting the number of contiguous 
spins with the same orientation.~\cite{maibaum2008phase}
In contrast,
our thermodynamic model does not rely on any clustering algorithm to identify nuclei of the different phases,
but instead only focuses on the total magnetization $M=\sum_i s_i$ 
as a macroscopic order parameter
to characterize the overall state
of the system.
The dotted blue line in Figure~\ref{fig:p5} indicates the free energy profile $\tilde{G}(M)$.
Taking a reference state with a 
circular Gibbs dividing surface, 
we computed the probability distribution $\rho\cond{M}{n_{up},n_{down}}$ from unbiased simulations of the bulk phases as shown in the Supplementary Material.
Using the iterative expression in
Eqn.~\eqref{eq:iterative},
and assuming the number of nucleation sites $N_s$ is the total size of the lattice, $L^2$,
we obtained $G(n_{up})$,
which corresponds to the free energy for a single positively-magnetized  cluster relative to the bulk negatively-magnetized phase.
$G(n_{up})$ is plotted as the solid red line in
Figure~\ref{fig:p5},
together with the free energy profile
$\tilde{G}(M)$ reconstructed
using Eqn.~\eqref{eq:gphigns}, that 
matches perfectly the directly computed free-energy
curve, signaling the convergence of self-consistent iterations.

In Ref.~\cite{ryu2010validity},
the computed nucleation free energy profile of the 2D Ising model was found to agree well with the expression
\begin{equation}
    G(n)=2\sqrt{\pi n} \sigma + \tau k_B T \ln n +d,
    \label{eq:cntwei}
\end{equation}
where $\sigma=1.20585$ is the temperature-dependent interfacial free energy for this Ising model that can be computed analytically~\cite{shneidman1999applicability},
$\tau k_B T \ln n$ accounts for the shape fluctuations of the cluster ($\tau=\dfrac{5}{4}$ for the 2D Ising model),
and the term $d=8-2\sqrt{\pi}\sigma$ ensures that the free energy of a isolated spin is correctly captured.
In Figure~\ref{fig:p5}, we plotted the exact prediction of Eqn.~\eqref{eq:cntwei} as the black dashed line. 
This prediction matches perfectly
the $G(n_{up})$ obtained from our thermodynamic framework, without
using any fitting parameters and 
without explicitly performing a 
cluster analysis of the simulation.

\begin{figure}
\includegraphics[width=0.5\textwidth]{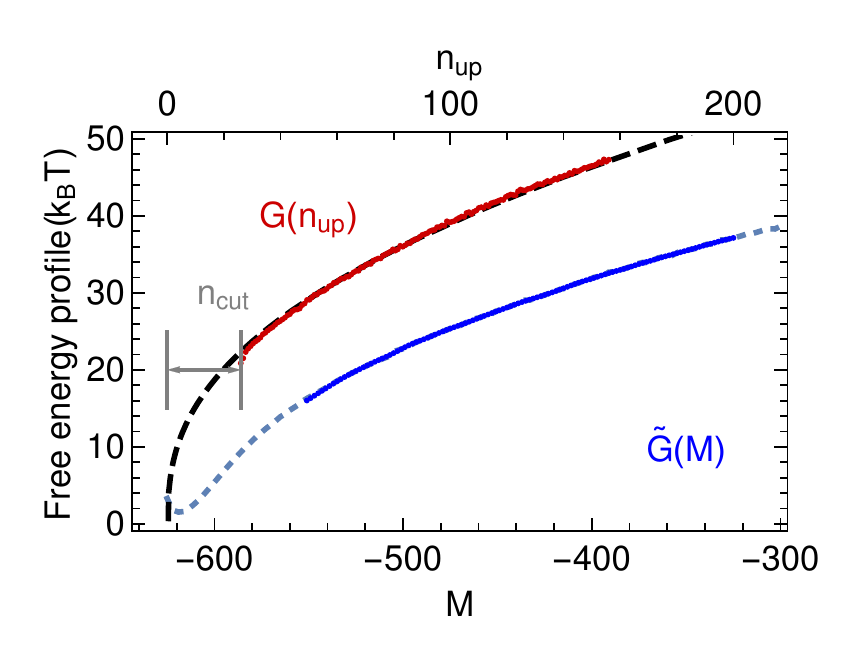}
\caption{
The dotted blue line is the free energy profile $\tilde{G}(M)$ as
a function of the magnetization, for a square-lattice Ising model with size $L=25$ and periodic boundary
conditions, simulated at $T=1.5$.
The solid red and the blue lines indicate the reconstructed 
curves for $G(n_{up})$ and $\tilde{G}(M)$,
respectively.
The dashed black line is the prediction from Eqn.~\eqref{eq:cntwei}.
Note that $G(n_{up})$ and $\tilde{G}(M)$ are discrete functions, as $n_{up}$ can only have integer values and $M$ can only be odd integers in this system.
The grey vertical lines indicate the choice of $\ncut$
for this system.
}
\label{fig:p5}
\end{figure}
}
\section{Conclusions}

The thermodynamic framework introduced in this paper 
provides a link between the molecular and the 
macroscopic scales. 
Any extensive quantity can be chosen to
discriminate between the solid and the liquid,
be it built upon local descriptors, 
or a traditional thermodynamic quantity 
such as the total volume, the energy
or the magnetization.
By characterizing the fluctuations
of this extensive quantity we can
rigorously define, in a probabilistic manner (Eqn.~\ref{eq:rhosl}),
{a reference state consistent with}
a zero-excess Gibbs dividing surface that 
encloses a single cluster of the stable phase.

Our method is applicable to all types of phase
transitions -- from solidification, to 
precipitation or condensation -- and it can be combined
with any sampling method one chooses
to accelerate nucleation~\cite{ten1996numerical,ten1999homogeneous,auer2001prediction,moroni2005interplay,trudu2006freezing,maibaum2008phase,lechner2011role,prestipino2012systematic,yi2012molecular,salvalaglio2015molecular,mccarty2016bespoke,piaggi2016fdreact16}.
From such simulations the 
free-energy for the overall system as a 
function of any extensive quantity can
be computed, 
and then converted into the free 
energy of a single cluster relative to the 
metastable bulk. By avoiding the need 
of singling out atom-size clusters that are 
inherently ill-defined, our approach is both
practically simple and conceptually
elegant. 
Since no assumption is made on the functional form 
of the computed free energy profile for nucleation, 
our approach can be used to test the limits
of classical nucleation theory, 
and extended so that it also describes heterogeneous 
nucleation, and thus further advances our understanding 
of bulk and interface-driven phase transitions.

\section{Supplementary Material}

The supplementary material contains a more
in-depth discussion of detailed procedures of our 
approach, together with descriptions of the
simulation protocols and a commented sample input file. 

\begin{acknowledgements}
The authors would like to thank Gareth Tribello and Gabriele Tocci
for insightful discussions and helpful comments,
and Massimiliano Bonomi for sharing with us data from well-tempered metadynamics simulations of the Ising model~\cite{bono-parr10prl}.
We also acknowledge funding from the Swiss National Science 
Foundation (Project ID 200021-159896).
\end{acknowledgements}

\foreach \x in {1,2,3,4,5,6}
{%
\clearpage


\section*{Supplementary material}

# Bridging the gap between atomistic and macroscopic models of homogeneous nucleation
## *Supplementary Material*

Bingqing Cheng[1, *] and Michele Ceriotti[1]

[1]*Laboratory of Computational Science and Modeling, Institute of Materials, École Polytechnique Fédérale de Lausanne, 1015 Lausanne, Switzerland*
(Dated: December 30, 2016)

## CLUSTER DISTRIBUTIONS UNDER THE MULTIPLE CLUSTER MODEL

In the multiple cluster model described in the main text, the probability of cluster formation follows

$$\langle p_n \rangle \propto \exp(-\beta G(n)), \tag{1}$$

with

$$G(n) = \mu n + \sigma v^{\frac{2}{3}} n^{\frac{2}{3}}, \tag{2}$$

and the effective interfacial free energy $\sigma$ is equal to the specific interfacial free energy $\gamma$ times a geometrical constant $\Omega$. For our toy system, we selected parameters $k_B = 1$, $T = 0.58$, $\mu = -0.0562\epsilon$, $\gamma = 0.354\epsilon/\sigma^2$, $\Omega = 4.836$, $v^{\frac{2}{3}} = 1.035\sigma^2$, and $\epsilon$ and $\sigma$ are the Lennard-Jones energy and length units. These parameters were selected to mimic those obtained in the calculations we performed for the Lennard-Jones system [1].

The probability distribution of the number of clusters that has $n$ atoms (denoted by $k_n$) are approximated by Poisson distributions, i.e.

$$\begin{aligned} P(n, k_n) &= \lambda(n)^{k_n} \exp(-\lambda(n))/k_n!, \\ \lambda(n) &= N_s \exp(-\beta G(n)), \end{aligned} \tag{3}$$

and we assumed $N_s = 700$.

The free energy profile $\tilde{G}(n_{\text{tot}})$ as a function of total number of atoms in all the clusters can be computed by enumerating all the possible combinations such as

$$\exp\left(-\beta \tilde{G}(n_{\text{tot}})\right) = \sum_{k_1=0}^{\infty} \sum_{k_2=0}^{\infty} \ldots \sum_{k_{n^\star}=0}^{\infty} \delta\left(\sum_{n=1}^{\infty} n k_n - n_{\text{tot}}\right) \prod_{n=1}^{n^\star} P(n, k_n), \tag{4}$$

where $\delta(x=0) = 1$ and $\delta(x \neq 0) = 0$, and $n^\star$ is the size of the critical nucleus. If a cutoff size $n_{\text{cut}}$ is selected, one can also compute the distribution of $n_{\text{tot}}$ under the constraint that no cluster in the system is of size larger than $n_{\text{cut}}$, i.e.

$$\exp\left(-\beta \tilde{G}^{(0)}(n_{\text{tot}})\right) = \sum_{k_1=0}^{\infty} \sum_{k_2=0}^{\infty} \ldots \sum_{k_{n_{\text{cut}}}=0}^{\infty} \delta\left(\sum_{n=1}^{\infty} n k_n - n_{\text{tot}}\right) \prod_{n=1}^{n_{\text{cut}}} P(n, k_n), \tag{5}$$

On the other hand, if one restricts the system to have exactly one cluster larger than $n_{\text{cut}}$, the distribution of $n_{\text{tot}}$ can be written as

$$\exp\left(-\beta \tilde{G}^{(1)}(n_{\text{tot}})\right) = \sum_{n_b=n_{\text{cut}}}^{n_{\text{tot}}} P(n_b, 1) \exp(-\beta \tilde{G}^{(0)}(n_{\text{tot}} - n_b)) \tag{6}$$

Furthermore, if one restricts the system to have two clusters that have sizes larger than $n_{\text{cut}}$, the distribution of $n_{\text{tot}}$ can be written as

$$\exp\left(-\beta \tilde{G}^{(2)}(n_{\text{tot}})\right) = \sum_{n_a=n_{\text{cut}}}^{n_{\text{tot}}} \sum_{n_b=n_{\text{cut}}}^{n_a} P(n_a, 1) P(n_b, 1) \exp(-\beta \tilde{G}^{(0)}(n_{\text{tot}} - n_a - n_b)) \tag{7}$$

We selected $n_{\text{cut}} = 150$, computed $\tilde{G}(n_{\text{tot}})$, $\tilde{G}^{(0)}(n_{\text{tot}})$, $\tilde{G}^{(1)}(n_{\text{tot}})$, and $\tilde{G}^{(2)}(n_{\text{tot}})$, and plotted them in Figure 1. It can be seen that at $n_{\text{tot}} > n_{\text{cut}} + \langle n_{\text{tot}} \rangle$, $\tilde{G}^{(1)}(n_{\text{tot}})$ overlaps with $\tilde{G}(n_{\text{tot}})$. In addition, at $n_{\text{tot}} > n_{\text{cut}} + \langle n_{\text{tot}} \rangle$ it is many orders of magnitude less likely to have zero or two clusters of size larger than $n_{\text{cut}}$ compared with only having one such cluster. Therefore, at $n_{\text{tot}} > n_{\text{cut}} + \langle n_{\text{tot}} \rangle$ the system effectively only contains one large cluster with size larger than $n_{\text{cut}}$, and $\tilde{G}^{(1)}(n_{\text{tot}})$ is a good approximation of $\tilde{G}(n_{\text{tot}})$. Note that $n_{\text{cut}}$ does not need to be decided exactly. As long as $n_{\text{cut}}$ is reasonably large, the conclusions above apply.

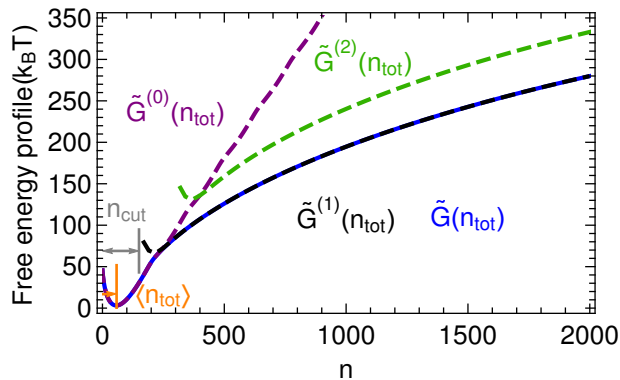

FIG. 1: Free energy profiles $\tilde{G}(n_{\text{tot}})$, $\tilde{G}^{(0)}(n_{\text{tot}})$, $\tilde{G}^{(1)}(n_{\text{tot}})$, and $\tilde{G}^{(2)}(n_{\text{tot}})$ of the atomistic system of multiple clusters. At $n_{\text{tot}} < n_{\text{cut}}$, $\tilde{G}^{(0)}(n_{\text{tot}})$ is exactly equal to $\tilde{G}(n_{\text{tot}})$. At $n_{\text{tot}} > n_{\text{cut}} + \langle n_{\text{tot}} \rangle$, $\tilde{G}^{(1)}(n_{\text{tot}})$ lies on the top of $\tilde{G}(n_{\text{tot}})$.

## THE CONNECTION BETWEEN THE MULTIPLE CLUSTER MODEL AND THE THERMODYNAMIC MODEL

In the main text, we introduced a thermodynamic model to describe the metastable liquid system, and argued that for sufficiently large $\Phi$ (here for simplicity we assume that the average order parameter in the solid phase is larger than in the liquid), the probability distribution of $\Phi$ in the liquid should follow

$$\rho_l(\Phi|N) = \int_{n_{\text{cut}}}^{n^\star} dn \rho(\Phi|n, N-n) N_s \exp(-\beta G(n)). \tag{8}$$

On the other hand, we also employed the multiple cluster model, and suggested that at $n_{\text{tot}} \gg n_{\text{cut}}$,

$$\exp(-\beta \tilde{G}(n_{\text{tot}})) = \sum_{n_b=n_{\text{cut}}}^{n_{\text{tot}}} N_s \exp(-\beta G(n)) \exp(-\beta \tilde{G}(n_{\text{tot}} - n_b)). \tag{9}$$

It is instructive to highlight the connection between Eqn. (8) of the thermodynamic model and Eqn. (9) of the multiple cluster model. By inserting the definition for $\rho(\Phi|n, N-n)$, and using the approximation $\rho_l(\Phi|N-n) \approx \rho_l(\Phi|N)$, (which can be regarded as equivalent to the assumption that the presence of other clusters do not affect the $P(n, k_n)$ in the multiple cluster model), one obtains

$$\rho_l(\Phi|N) = \int d\varphi \rho_l(\Phi - \varphi|N) N_s \exp(-\beta G(\varphi)), \tag{10}$$

where we defined the free energy for a cluster characterized by an order parameter $\varphi$,

$$\exp(-\beta G(\varphi)) = \int_{n_{\text{cut}}}^{n^\star} dn \rho_s(\varphi|n) \exp(-\beta G(n)). \tag{11}$$

Eqn. (10) corresponds to a continuous version of Eqn. (9), in which the extensive order parameter $\Phi$ is used to characterize the solid fraction in the metastable liquid and the size of the largest solid cluster.

## CONDITIONAL PROBABILITY DISTRIBUTIONS $\rho_s(\Phi|n)$ AND $\rho_l(\Phi|n)$ IN THE LENNARD-JONES SYSTEM

For the Lennard-Jones system described in the main text, assuming the Gibbs dividing surface separating the solid cluster and the bulk liquid has a spherical shape, we can compute the two probability distributions $\rho_s(\Phi|n)$ and $\rho_l(\Phi|n)$ from simulations of the bulk phases.

For example, take a molecular dynamics simulation run of a bulk solid system that has $N$ atoms. From each snapshot, one can randomly select a volume that has a spherical shape, as illustrated in Figure 2. One can then count the number of atoms $n$ in that volume, as well as the extensive quantity $\Phi$ associated with those $n$ atoms. After accumulating enough statistics for $(n_i, \Phi_i)$, one can build the conditional probability distribution $\rho_s(\Phi|n)$ by

$$\rho_s(\Phi|n) = \langle \delta(\Phi_i - \Phi)\delta(n_i - n)\rangle / \langle \delta(n_i - n)\rangle, \tag{12}$$

2

3

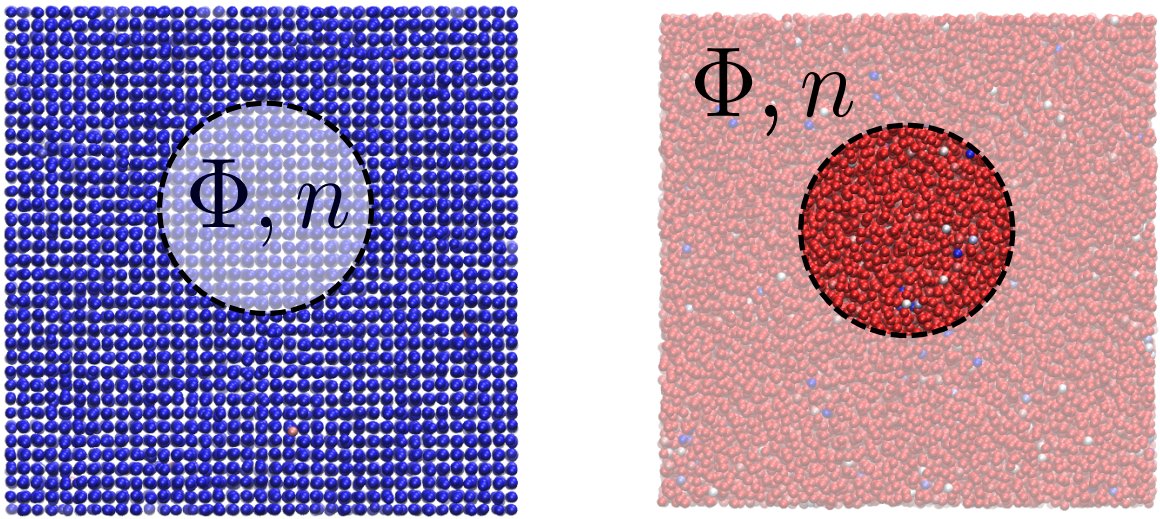

FIG. 2: A schematic illustration of selecting a contiguous portion of $n$ atoms in a system.

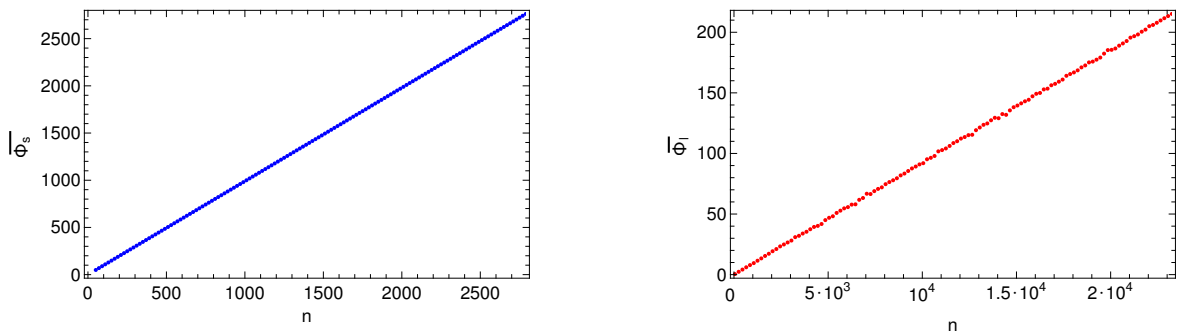

FIG. 3: The average extensive quantity $\Phi$ for different system sizes.

where $\langle \ldots \rangle$ denotes the ensemble average at the given thermodynamic condition. Furthermore, $\rho_s(\Phi|n)$ can also be computed from biased simulations, by statistical re-weighting

$$\rho_s(\Phi|n) = \langle \delta(\Phi_i - \Phi)\delta(n_i - n)\exp(\beta V_i)\rangle / \langle \delta(n_i - n)\exp(\beta V_i)\rangle, \tag{13}$$

where $V_i$ is the instantaneous external bias for each snapshot.

For the bulk solid system, the average extensive quantity

$$\overline{\Phi_s}(n) = \int \mathrm{d}\Phi \rho_s(\Phi|n) \Phi \tag{14}$$

at each $n$ was computed and plotted in the left panel of Figure 3. It can be seen that scales linearly with system size, and thus can be expressed as $\overline{\Phi_s} = n\phi_s$, where $\phi_s$ is the average value of order parameter of each atom in the bulk solid. The average $\Phi$ in bulk liquid $\overline{\Phi_l}(n) = \int \mathrm{d}\Phi \rho_l(\Phi|n)$ is also shown in the right panel of Figure 3. Similarly, in bulk liquid $\overline{\Phi_l} = n\phi_l$, where $\phi_l$ is the average value of order parameter of each atom in the bulk solid. From simulations we determined $\phi_s = 0.99000$ and $\phi_l = 0.00936$. We have also computed the variances of $\Phi$ in the bulk solid and the bulk liquid at different $n$, using the equations

$$\mathrm{Var}(\Phi_s) = \int \mathrm{d}\Phi (\Phi - n\phi_s)^2 \rho_s(\Phi|n) \tag{15}$$

and

$$\mathrm{Var}(\Phi_l) = \int \mathrm{d}\Phi (\Phi - n\phi_l)^2 \rho_l(\Phi|n). \tag{16}$$

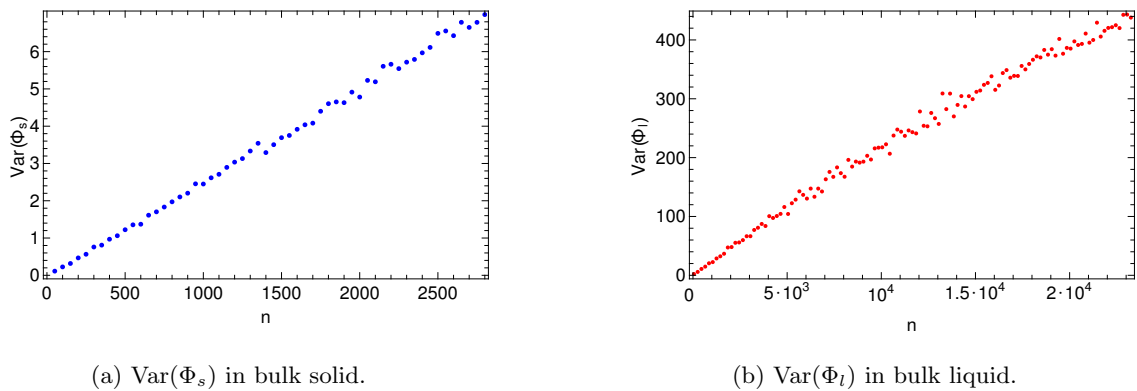

(a) Var($\Phi_s$) in bulk solid.   (b) Var($\Phi_l$) in bulk liquid.

FIG. 4: The variance of the extensive quantity $\Phi$ at different sizes.

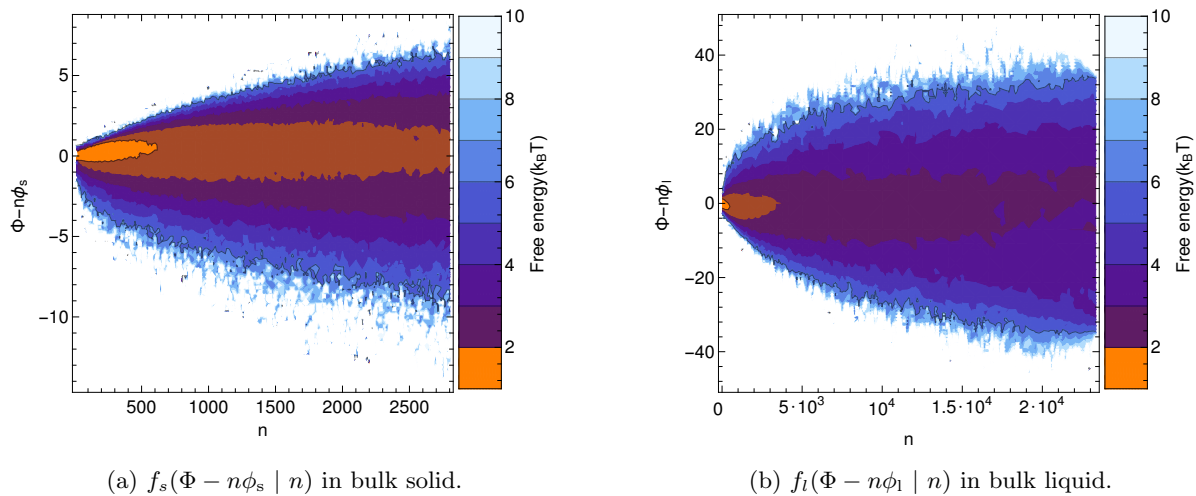

(a) $f_s(\Phi - n\phi_s \mid n)$ in bulk solid.   (b) $f_l(\Phi - n\phi_l \mid n)$ in bulk liquid.

FIG. 5: The free energies associated with the quantities $\Phi - n\phi_s$ and $\Phi - n\phi_l$ at different system size.

The computed variances of $\Phi$ are plotted in Figure 4, which shows that the variances of $\Phi$ in both phases also scale linearly with the number of the atoms. Furthermore, in Figure 5 we plot the quantities $f_s(\Phi - n\phi_s \mid n)$ and $f_l(\Phi - n\phi_l \mid n)$, which are defined as

$$f_s(\Phi - n\phi_s \mid n) = -\log(\rho_s(\Phi \mid n)), \tag{17}$$

and

$$f_l(\Phi - n\phi_l \mid n) = -\log(\rho_l(\Phi \mid n)). \tag{18}$$

The parabolic shapes of $f_s(\Phi - n\phi_s \mid n)$ and $f_l(\Phi - n\phi_l \mid n)$ suggest that $\rho_s(\Phi|n)$ and $\rho_l(\Phi|n)$ follow Gaussian-like distributions centered at $n\phi_s$ and $n\phi_l$, respectively. All in all, the analyses above suggest that $\rho_s(\Phi|n)$ and $\rho_l(\Phi|n)$ can be approximated as the distributions of the sum of $n$ independent Gaussian-distributed random variables.

## THE PROBABILITY DISTRIBUTION $\rho(M|n_{up}, n_{down})$ OF SQUARE-LATTICE ISING MODEL

For the square-lattice Ising model with $L^2$ lattice described in the main text, the probability distribution $\rho(M|n_{up}, n_{down})$ can be determined as

$$\rho(M|n_{up}, n_{down}) = \int dM' \rho_{up}(M'|n_{up}) \rho_{down}(M - M'|n_{down}). \tag{19}$$

In general $\rho_{up}(M|n) = \rho_{down}(-M|n)$ because of the symmetry in the system, therefore only one of them needs to be characterized.

We took a Monte Carlo simulation of the Ising lattice, and from each snapshot we can randomly selected a circle and then counted the number of up spins $n$ as well as the magnetization $M$ inside ( and outside) the circle.

5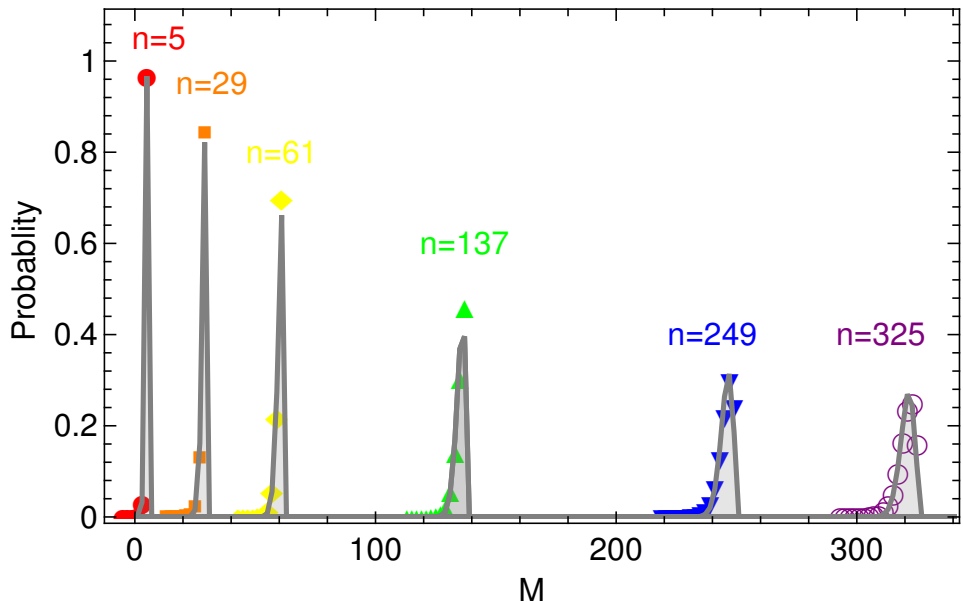

FIG. 6: The probability distributions $\rho_{up}(M|n)$ at different values of $n$. The gray lines indicate the predictions from the Binomial model.

Whether the $n$ lattices were selected inside or outside the circle was found to have negligible effect. We plot some of the probability distributions $\rho_{up}(M|n)$ at different values of $n$ in Figure 6.

We also found a Binomial model that is able to describe the behavior of $\rho_{up}(M|n)$ very well at all values of $n$. In this Bionomial model, we assume each spin has a probability of $p = 0.00674$ to flip to the opposite site in the sea of up spins. As such, $\rho_{up}(M|n)$ can be approximated as

$$\rho_{up}(M|n) \approx \binom{n}{k} p^k (1-p)^{n-k}, \ k = (M-n)/2. \tag{20}$$

In Figure 6, we also plotted the predictions of this model, which find good agreement with the distributions computed directly from the simulations.

### Sample LAMMPS input file

Here is a sample LAMMPS input file for running the molecular dynamics simulations.

```
atom_style     atomic
units          lj
dimension      3
boundary       p p p
processors     * * 1

read_data      ./data.lj    # Coordinate file
```

Atoms of type 1 are the Lennard-Jones particles with the truncated pairwise potential [2–4]. The atom of type 2 is a fixed ghost atom that does not interact with its surroundings. It is used as reference point in some of the collective variables.

```
group real type 1
velocity real create 0.6 RandomSeedHere dist gaussian

pair_style table linear 5000
pair_coeff 1 1 tr-lj.table TR_LJ 2.5
pair_coeff * 2 tr-lj.table NULL 2.5

neighbor        1.0 bin

timestep 0.004
```

An initial equilibration of 20'000 steps is performed following by a complete melting of the system. During these processes a highly efficient Generalized Langevin thermostat [5] is employed.

```
fix 1 real press/berendsen iso 0.0 0.0 0.5
fix 2 real gle 6 0.84 0.84 12rrr5 smart2.A every 2
run 80000
unfix 2
fix 3 real gle 6 0.58 0.58 12rrr5 smart2.A every 5
run 20000
unfix 3
```

After the equilibration a metadynamics simulation in the NPT ensemble, using a Nosé-Hoover barostat and thermostat is started.

```
fix 4 real npt temp 0.58 0.58 0.2 iso 0.0 0.0 0.5
fix 5 all plumed   plumedfile plumed.dat    outfile p.log

thermo 500
thermo_style one
dump coord all xyz 5000 traj.xyz

run             20000000
write_restart restart.lj
```

## Sample PLUMED input file

Here we provide a sample input file for PLUMED. This input was used to compute the free energy profile during homogenous nucleation.

```
UNITS NATURAL
RESTART
```

Record the volume of the simulation cell.

```
cell: CELL
```

We first compute the local order parameter $\kappa$ using the cubic harmonic function "FCCUBIC". The sigmoid switching function "SMAP" is used to do a non-linear mapping on $\kappa$. We collect $\Phi = \sum S(\kappa(i))$ - the global order parameter used to identify $n_s^\Phi$ - using the keyword "MORE_THAN". The value of "cub.morethan" is proportional to the number of atoms in the *fcc* crystal.
The system contains a total of 23328 real atoms, The first atom is a "ghost" atom that stays stationary at the center of the supercell. This ghost atom is only used as a reference point.

```
FCCUBIC ...
LABEL=cub
SPECIES=2-23329 SWITCH={CUBIC D_0=1.2 D_MAX=1.5} ALPHA=27 MEAN
MORE_THAN1={SMAP R_0=0.45 D_0=0.0 A=8 B=8}
... FCCUBIC
```

Finally the instantaneous values of the CV and the biases are recorded. These are used in the re-weighting processes to construct the FES, e.g.

$$G(\Phi) = -\frac{1}{\beta} ln \left( \frac{\int dt e^{\beta V_{tot}(t)} \delta(\Phi)}{\int dt e^{\beta V_{tot}(t)}} \right), \tag{21}$$

where $\beta = 1/k_B T$, and $V_{tot}(t)$ is the total external bias including the metadynamics bias and the restraint biases at time $t$.

```
METAD ...
LABEL=metad
ARG=cub.morethan-1
PACE=1000 HEIGHT=0.3 SIGMA=600 FILE=HILLS
TEMP=0.58 BIASFACTOR=400
ADAPTIVE=DIFF SIGMA_MAX=300 SIGMA_MIN=0.1
... METAD

fixmax: UPPER_WALLS ARG=cub.morethan-1 AT=2500 KAPPA=0.1

# monitor the two variables and the metadynamics bias potential
PRINT STRIDE=10 ARG=cell.ax,cub.*,metad.bias,fixmax.bias FILE=COLVAR

ENDPLUMED
```


}
\end{document}